\input harvmac

\skip0=\baselineskip
\divide\skip0 by 2
\def\tmpsp{\the\skip0}

\let\linesp=\mylinesp

\def\itwo{\textstyle{i \over 2}}
\def\CN{{\cal N}}
\def\p{\partial}
\def\bx{{\bf x}}
\def\cQ{{\cal{Q}}}
\def\cD{{\cal{D}}}

\lref\twoBH{
R.~Britto-Pacumio, A.~Strominger and A.~Volovich,
``Two-black-hole bound states,''
JHEP {\bf 0103}, 050 (2001)
[arXiv:hep-th/0004017].
}

\lref\george{
G.~Papadopoulos,
``Conformal and superconformal mechanics,''
Class.\ Quant.\ Grav.\  {\bf 17}, 3715 (2000)
[arXiv:hep-th/0002007].
}

\lref\sqm{
J.~Michelson and A.~Strominger,
``The geometry of (super)conformal quantum mechanics,''
Commun.\ Math.\ Phys.\  {\bf 213}, 1 (2000)
[arXiv:hep-th/9907191].
}

\lref\coles{
R.~A.~Coles and G.~Papadopoulos,
``The Geometry Of The One-Dimensional Supersymmetric Nonlinear Sigma Models,''
Class.\ Quant.\ Grav.\  {\bf 7}, 427 (1990).
}

\lref\jackiw{
R.~Jackiw,
``Dynamical Symmetry Of The Magnetic Monopole,''
Annals Phys.\  {\bf 129}, 183 (1980).
}

\lref\cj{
S.~R.~Coleman and R.~Jackiw,
``Why Dilatation Generators Do Not Generate Dilatations?,''
Annals Phys.\  {\bf 67}, 552 (1971).
}

\lref\ccj{
C.~G.~Callan, S.~R.~Coleman and R.~Jackiw,
``A New Improved Energy - Momentum Tensor,''
Annals Phys.\  {\bf 59}, 42 (1970).
}

\lref\gt{
G.~W.~Gibbons and P.~K.~Townsend,
``Black holes and Calogero models,''
Phys.\ Lett.\ B {\bf 454}, 187 (1999)
[arXiv:hep-th/9812034].
}

\lref\dictionary{
L.~Frappat, P.~Sorba and A.~Sciarrino,
``Dictionary on Lie superalgebras,''
arXiv:hep-th/9607161.
}

\lref\ddf{
V.~de Alfaro, S.~Fubini and G.~Furlan,
``Conformal Invariance In Quantum Mechanics,''
Nuovo Cim.\ A {\bf 34}, 569 (1976).
}

\lref\ferrell{
R.~C.~Ferrell and D.~M.~Eardley,
``Slow Motion Scattering And Coalescence Of Maximally Charged Black Holes,''
Phys.\ Rev.\ Lett.\  {\bf 59}, 1617 (1987).
}

\lref\fived{
J.~Michelson and A.~Strominger,
``Superconformal multi-black hole quantum mechanics,''
JHEP {\bf 9909}, 005 (1999)
[arXiv:hep-th/9908044].
}

\lref\hktokt{
G.~W.~Gibbons, G.~Papadopoulos and K.~S.~Stelle,
``HKT and OKT geometries on soliton black hole moduli spaces,''
Nucl.\ Phys.\ B {\bf 508}, 623 (1997)
[arXiv:hep-th/9706207].
}

\lref\hull{
C.~M.~Hull,
``The geometry of supersymmetric quantum mechanics,''
arXiv:hep-th/9910028.
}

\lref\iceland{
R.~Britto-Pacumio, J.~Michelson, A.~Strominger and A.~Volovich,
``Lectures on superconformal quantum mechanics and multi-black hole
moduli spaces,''
Proceedings of the NATO ASI on Quantum Geometry, Akureyri, Iceland (1999)
[arXiv:hep-th/9911066].
}

\lref\jerfour{
J.~Michelson,
``Scattering of four-dimensional black holes,''
Phys.\ Rev.\ D {\bf 57}, 1092 (1998)
[arXiv:hep-th/9708091].
}

\lref\ft{
J.~Traschen and R.~Ferrell,
``Quantum mechanical scattering of charged black holes,''
Phys.\ Rev.\ D {\bf 45}, 2628 (1992)
[arXiv:hep-th/9205061].
}

\lref\shiraishi{
K.~Shiraishi,
``Moduli space metric for maximally charged dilaton black holes,''
Nucl.\ Phys.\ B {\bf 402}, 399 (1993).
}

\lref\gk{
G.~W.~Gibbons and R.~E.~Kallosh,
``Topology, Entropy And Witten Index Of Dilaton Black Holes,''
Phys.\ Rev.\ D {\bf 51}, 2839 (1995)
[arXiv:hep-th/9407118].
}

\lref\azcarraga{
J.~A.~de Azcarraga, J.~M.~Izquierdo, J.~C.~Perez Bueno and P.~K.~Townsend,
``Superconformal mechanics and nonlinear realizations,''
Phys.\ Rev.\ D {\bf 59}, 084015 (1999)
[arXiv:hep-th/9810230].
}

\lref\kumar{
J.~Kumar,
``Conformal mechanics and the Virasoro algebra,''
JHEP {\bf 9904}, 006 (1999)
[arXiv:hep-th/9901139].
}

\lref\gpat{
G.~Papadopoulos and A.~Teschendorff,
``Multi-angle five-brane intersections,''
Phys.\ Lett.\ B {\bf 443}, 159 (1998)
[arXiv:hep-th/9806191].
}

\lref\gpjg{
J.~Gutowski and G.~Papadopoulos,
``The dynamics of very special black holes,''
Phys.\ Lett.\ B {\bf 472}, 45 (2000)
[arXiv:hep-th/9910022].
}

\lref\gibb{
G.~W.~Gibbons and P.~J.~Ruback,
``The Motion Of Extreme Reissner-Nordstrom Black
Holes In The Low Velocity Limit,''
Phys.\ Rev.\ Lett.\  {\bf 57}, 1492 (1986).
}

\lref\nahm{
W.~Nahm,
``Supersymmetries And Their Representations,''
Nucl.\ Phys.\ B {\bf 135}, 149 (1978).
}

\lref\parker{
M.~Parker,
``Classification Of Real Simple Lie Superalgebras Of Classical Type,''
J.\ Math.\ Phys.\  {\bf 21}, 689 (1980).
}

\lref\wyllard{
N.~Wyllard,
``(Super)conformal many-body quantum mechanics with extended  supersymmetry,''
J.\ Math.\ Phys.\  {\bf 41}, 2826 (2000)
[arXiv:hep-th/9910160].
}

\lref\kallosh{
P.~Claus, R.~Kallosh and A.~Van Proeyen,
``Conformal symmetry on world volumes of branes,''
Proceedings of the Trieste Conference on Superfivebranes and Physics
in $5+1$ dimensions (1998)
[arXiv:hep-th/9812066].
}

\lref\claus{
P.~Claus, M.~Derix, R.~Kallosh, J.~Kumar, P.~K.~Townsend and A.~Van Proeyen,
``Black holes and superconformal mechanics,''
Phys.\ Rev.\ Lett.\  {\bf 81}, 4553 (1998)
[arXiv:hep-th/9804177].
}

\Title{\vbox{\baselineskip12pt
        \hbox{HUTP-99/A055}
        \hbox{hep-th/9911001}
}}{\vbox{\centerline{Superconformal Multi-Black Hole Moduli Spaces}
\vskip 0.2cm
\centerline
{in Four Dimensions}}}

\centerline{
        Alexander Maloney,
        Marcus Spradlin
        and Andrew Strominger
}

\centerline{Department of Physics}
\centerline{Harvard University}
\centerline{Cambridge, MA 02138}

\vskip .3in
\centerline{\bf Abstract}

Quantum mechanics on the moduli space of $N$ supersymmetric 
Reissner-Nordstrom black
holes is shown to admit
4 supersymmetries using an unconventional supermultiplet which
contains $3N$ bosons and $4N$ fermions. A near-horizon
limit is found in which the quantum mechanics of widely separated black holes
decouples from that of strongly-interacting, near-coincident black
holes. This near-horizon theory is shown to have an enhanced $D(2,1;0)$
superconformal symmetry. The bosonic symmetries are $SL(2,{\bf R})$ conformal 
symmetry and $SU(2)\times SU(2)$ $R$-symmetry arising from 
spatial rotations and the $R$-symmetry of $\CN = 2$ supergravity. 

\smallskip

\Date{November 1999}

\listtoc
\writetoc

\newsec{Introduction}

The quantum mechanics of $N$ slowly-moving, four-dimensional
extremal Reissner-Nordstrom black holes is a sigma model whose
target space is the moduli space of multi-black hole solutions.
This moduli space is parameterized by the $3N$ coordinates of the
$N$ black holes in ${\bf R}^3$.  The metric on this moduli space was
discovered over a decade ago by Ferrell and Eardley \refs{\ferrell,\gibb}.
When embedded in $\CN=2$ supergravity, the static black hole
configurations preserve four of the eight supersymmetries. One
therefore expects an $\CN=4$ supersymmetric quantum mechanics 
with $4N$ fermions arising as goldstinos.
Typically four supersymmetries would imply a complex structure on
the target space. This is clearly not possible in this instance,
however, because the dimension of the moduli space can be odd. This
puzzle is discussed in \refs{\gk \hktokt-\gpat}. Aspects of this
problem are explored in \refs{\ft \shiraishi - \jerfour}.

In this paper we resolve this puzzle by coupling $3N$ real $\CN=1$
supermultiplets $\Phi^\mu$, containing one fermion and one boson
apiece, to $N$ real fermionic $\CN=1$ supermultiplets $\psi^A$,
each containing one physical fermion and an auxiliary boson.
This gives the required $3N$
bosons and $4N$ fermions. We show that taken together these comprise a
constrained $\CN=4$ multiplet, which is then used to construct a
general class of $\CN=4$ actions.  The geometry of such theories
is a generalization of the weak hyperk\"ahler
with torsion geometry
\hktokt\ to $3N$
dimensions, in which the $SU(2)$ generators of ${\bf R}^3$ spatial rotations
play the role of the triplet of complex structures.
The  Ferrell-Eardley moduli space is shown to be an example of  
such a  geometry and therefore admits $\CN=4$ supersymmetry, although
we will see that the moduli space metric is modified when the
auxiliary bosons are integrated out.
Supersymmetry therefore requires corrections to the Ferrell-Eardley
metric.

We further consider a near-coincident or near-horizon limit of the
moduli space in which the coordinate separation (in spatially conformal
coordinates) of the center of
mass of the black holes is small compared to their size. 
In this
limit the actual geodesic distance between horizons remains
infinite and all curvatures remain small, so the semiclassical
approximation is expected to be valid. 
At low energies  this near-horizon quantum mechanics completely decouples from
quantum mechanics of widely-separated black holes. We show that
the near-horizon theory has an enhanced $D(2,1;0)$ superconformal
symmetry. One of the bosonic $SU(2)$ subgroups of $D(2,1;0)$
arises from spacetime rotations, while the other arises form the
$R$-symmetry of $\CN=2$ supergravity in four dimensions.  

One motivation for this work is to understand the spectrum of 
black hole bound states. The wave function for any state 
of the near-horizon theory 
has coordinate separations between black holes which are 
small compared to their
size. Such states are therefore 
multi-black hole bound states. We expect that the $D(2,1;0)$ superconformal
symmetry will play a key role in understanding the bound state
spectrum.

In section 2 we describe the moduli space and its near-horizon
limit.  In section 3 we construct the $\CN=4$ supersymmetric
extension.  In section $4$ we describe $D(2,1;\alpha)$
superconformal quantum mechanics.  In section 5 we show that the
near-horizon theory has $D(2,1;0)$ superconformal symmetry.
Related work in five dimensions appears in \refs{\sqm \fived
\gpjg - \iceland}.  Related work on supersymmetric and superconformal quantum 
mechanics appears in \refs{\gk,\hktokt,\azcarraga \kumar \hull - \wyllard}.
Some aspects of sections 3 and 4 have been investigated independently
by G. Papadopoulos \george.

\newsec{The Multi-Black Hole Moduli Space}

We wish to study the moduli space of extremal black hole solutions of
pure $\CN = 2$ supergravity in four dimensions.  To this end we will
first review results for the moduli space of black hole solutions in
Einstein-Maxwell theory, which is the bosonic sector of the
supergravity theory under consideration.

\subsec{The Moduli Space Metric}

The study of black hole moduli spaces was pioneered by Ferrell and
Eardley \ferrell, who considered extremally charged black holes in
four-dimensional Einstein-Maxwell theory, which has the action
\eqn\sry{S_{EM}={1 \over 16 \pi}\int d^4x\sqrt{-g}(R-F^2).}
This theory admits
static
multi-black hole solutions whose metric and potential are given by
\eqn\solution{
ds^2 = - \psi^{-2} dt^2 + \psi^2 d{\bf x}^2, ~~~~~
A = - (1 - \psi^{-1}) dt
}
in terms of a harmonic function
\eqn\psidef{
\psi({\bf x}) = 1 + \sum_A {m_A \over |{\bf x} - {\bf x}^A|},
}
where $A = 1,\ldots,N$ labels the $N$ black holes with masses $m_A$
and positions ${\bf x}^A$.  The Einstein-Maxwell action evaluated for
solution \solution
\eqn\aaa{
\CL_0 = - \sum_A m_A
}
is independent of ${\bf x}^A$, thus the positions of the black holes
are moduli.  This is the well known fact that the electric repulsion
and gravitational attraction of extremal black holes cancel.  If one
gives the black holes small velocities ${\bf v}^A$ and expands
the Einstein-Maxwell action to $\CO(v^2)$ around the static solution
\solution\ one finds Ferrell and Eardley's effective lagrangian
\ferrell
\eqn\eardley{
\CL = {1 \over 2} \sum_A m_A (v^A)^2 + \CL_{\rm int},
}
where
\eqn\interaction{
\CL_{\rm int} = {3 \over 8 \pi} \int d^3 x \ \psi^2 \sum_{A,B}
{m_A m_B \over r_A^3 r_B^3} \left[ \half ({\bf r}_A \cdot {\bf r}_B)
|{\bf v}^A - {\bf v}^B|^2 - ({\bf r}_A \times {\bf r}_B) \cdot
({\bf v}^A \times {\bf v}^B)\right],
}
${\bf r}_A = {\bf x} - {\bf x}^A$ and $r_A = | {\bf r}_A|$.  It is a
curious (and unexplained) fact that \interaction\ contains only
two-body, three-body and four-body interactions.
A very useful form of the effective lagrangian \eardley\ is
\eqn\modmetric{
\CL = \ha \sum_{A,B} v^{Ak} v^{Bl} (\delta^i_k \delta^j_l
+ \epsilon^{mi}{}_k {\epsilon_m}{}^{j}{}_l) \p_{Ai} \p_{Bj} L,
\qquad
L = - {1 \over 16 \pi} \int d^3 x\ \psi^4,
}
with $\psi$ given by \psidef\ and spatial indices $i,j=1,2,3$.

\subsec{The Near-Horizon Limit}

Let us first consider the single black hole solutions of
Einstein-Maxwell theory. For an extremal black hole at ${\bf x}^1$
the metric takes the form \solution\ where $\psi({\bf x}) = 1 + {m
\over |{\bf x} - {\bf x}^1|}$. The near-horizon limit is defined
by $|{\bf x} - {\bf x}^1| \ll 1$, in which case the metric takes the
form \eqn\nearg{ ds^2 = - \left({r \over m}\right)^{2} dt^2 +
\left({m \over r}\right)^2 dr^2 + m^2 d\Omega^2 } where $r=|\bx -
\bx^1|$. Thus the near-horizon geometry of a single extremal black
hole is $AdS_2 \times S_2$.

This analysis of the near-horizon limit of the physical geometry
motivates the definition of a similar limit for the multi-black hole
moduli space geometry.
We require
that $|\bx^A - \bx^B| \ll 1$ for $A,B = 1,\dots,N$.
In this limit the 1 in the harmonic function $\psi$
can be dropped and the potential in \modmetric\ is replaced by 
\eqn\nearl{ L = -
{1 \over 16 \pi} \int d^3 x\ \left[\sum_A {m_A \over |{\bf x} -
{\bf x}^A|}\right]^4 
.} 
The geometry of the moduli space is still
quite complicated. An important feature of this geometry is the
existence of noncompact, asymptotically locally flat  regions for
${\bf x}^A \to  {\bf x}^B$. These correspond to near-coincident
black holes.

\newsec{$\CN = 4$ Supersymmetry and the Black Hole Moduli Space}

In this section we demonstrate that \modmetric\ admits an $\CN = 4$
supersymmetric extension.  Such an extension is expected because
the solution \solution\ preserves four supersymmetries when embedded
in $\CN=2$, $D=4$ supergravity.  The four broken supersymmetries
are expected to lead to four goldstinos per black hole,
and hence $4N$ fermions in addition to the $3N$ bosons in 
the supersymmetric quantum mechanics.  Since the unbroken supercharges 
transform under spatial rotations, we further expect an $SU(2)$ $R$ 
symmetry with the bosons in triplets and the fermions in doublets. 
A second $SU(2)$ quantum mechanical $R$-symmetry, with singlet bosons 
and doublet fermions, is expected from 
reduction of the four-dimensional $R$-symmetry.

This field content sounds rather exotic as one usually encounters
equal numbers of bosons and fermions in a supersymmetric theory. 
However in one dimension there is
a fermion supermultiplet $\Psi$ whose only physical
fields are fermions \coles.
Accordingly in this section we consider
$3N$ real $\CN = 1$ multiplets $\Phi^\mu$ -- each of which contains one 
boson and one fermion -- along with $N$ extra fermion
multiplets $\Psi^A$ -- each of which contains a single physical
fermion.  These are combined into a constrained $\CN=4$ 
multiplet in a  manner which properly realizes the $R$-symmetries.  
Invariant actions are then constructed using $\CN=4$ superspace. 
This construction is found to include the black hole quantum mechanics as a special
case. 

\subsec{Supersymmetry Transformations}

Our treatment of supersymmetry follows closely that of Coles
and Papadopoulos \coles, although our
notation differs very slightly.
We first introduce the $\CN=1$ superfields
\eqn\superfields{
\Phi^\mu = X^\mu - i \theta_{} \lambda^\mu, \qquad
\Psi^A = i \psi^A + i \theta_{} b^A,
}
where $X^\mu$ and $b^A$ are real bosons and $\lambda^\mu$ and
$\psi^A$ are real fermions.  The bosonic superfield $\Phi^\mu$
is the usual map from $\CN=1$ superspace into the sigma model
manifold ${\cal M}$.
We define the usual superspace derivatives for the $0^{th}$ supersymmetry
\eqn\qdef{
\cQ_{} = {d \over d\theta_{}} + i\theta_{} {d \over dt}, \qquad
\cD_{} = {d \over d\theta_{}} - i\theta_{} {d \over dt}
}
which obey
\eqn\qd{
\cQ_{}^2 = i {d \over dt}, \qquad \cD_{}^2 = -i {d \over dt}, \qquad
\{\cQ_{},\cD_{}\} = 0.
}
For the remaining supersymmetries, $i=1,\ldots,\CN-1$, one makes
the ansatz
\eqn\extenansa{
\cQ_i \Phi^\mu = I_i{}^\mu{}_\nu \cD \Phi^\nu +
e_i{}^\mu{}_A \Psi^A, \qquad
\cQ_i \Psi^A = I_i{}^A{}_B \cD \Psi^B - i  e_i{}^A{}_\mu
\dot{\Phi}^\mu,
}
where $\dot{} = {d \over dt}$.
Under an infinitesimal supersymmetry transformation parametrized
by $\zeta^r$, $r=0,\ldots,{\CN}-1$,
the superfields $\Phi^\mu$ and $\Psi^A$
transform as
\eqn\susyxform{
\delta_\zeta \Phi^\mu = \zeta^r \cQ_r \Phi^\mu, \qquad \delta_\zeta
\Psi^A = \zeta^r \cQ_r \Psi^A,
}
where $\cQ_0\equiv \cQ$. 
The conditions that \qdef\ and \extenansa\ close to the supersymmetry
algebra
\eqn\susyalgebra{
\{ \cQ_r, \cQ_s \} = 2 i  \delta_{rs} {d \over dt},
\qquad r,s = 0,\ldots,\CN-1,
}
appear in Appendix A of \coles\ (and in much more generality than
our special case).
The most interesting constraints are
\eqn\interest{
-I_i{}^\mu{}_\nu I_i{}^\nu{}_\rho + e_i{}^\mu{}_A e_i{}^A{}_\rho
= \delta^\mu_\rho, \qquad \CN(I_i)^\mu{}_{\nu \rho} = 0
}
for all $i$,
where $\CN(I_i)$ is the Nijenhuis tensor of $I_i$.

If the $e_i$ terms are absent \interest\ requires the $I_i$
to be complex structures.  This is impossible, however, if the
target space is $3N$ dimensional as in the case of current interest.
A simple solution of the constraints when $\mu=1,\dots,3N$ and
$A=1,\dots,N$ is
\eqn\ourie{
(I_i){}^{A j}{}_{B k} = \delta^A_B \epsilon_i{}^j{}_k, \qquad
(I_i){}^A{}_B = 0, \qquad
(e_i){}^A{}_{B j} = \delta^A_B \delta_{ij}, \qquad
(e_i){}^{B j}{}_A = \delta^B_A \delta_i^j,
}
where we have replaced the index $\mu$ with the index pair
$Ai$ with $i=1,2,3$. It is easy to check directly that the 
closure conditions \susyalgebra\ are satisfied by \ourie. 

\subsec{Supersymmetric Actions}

In order to construct supersymmetric actions it is efficient to
introduce constrained $\CN=4$ superfields.
We employ anticommuting superspace
coordinates $\theta_r$, $r=0,\dots,3$, where $\theta_0 \equiv \theta $
is the usual $\CN=1$ superspace coordinate. The
corresponding $\CN=4$ superfields are denoted in boldface 
${\bf \Phi}^{\mu} (t,\theta_r)$
and ${\bf \Psi}^A(t, \theta_r)$. Their $\theta_i$-independent components
are the usual $\CN=1$ superfields ${ \Phi}^{\mu} (t,\theta_{})$
and $ \Psi^A(t, \theta_{})$ appearing in \superfields.
We define
\eqn\ddef{
\cQ_r = {d \over d \theta_r} + i \theta_r {d \over dt}, \qquad
\cD_r = {d \over d \theta_r} - i \theta_r {d \over dt}
}
which obey
\eqn\extendobey{
\{ \cQ_r, \cQ_s \} = 2 i \delta_{rs} {d \over d t} = - \{ \cD_r, \cD_s \},
\qquad \{ \cQ_r, \cD_s \} = 0,
}
with $\cD_0\equiv \cD$.
The supersymmetry transformations in $\CN=4$ superspace are
generated by the $\cQ_r$,
\eqn\extendedtrans{
\delta_\zeta {\bf \Phi}^\mu = \zeta^r \cQ_r {\bf \Phi}^\mu, \qquad
\delta_\zeta {\bf \Psi}^A = \zeta^r \cQ_r {\bf \Psi}^A,
}
which automatically obey the supersymmetry algebra.
The $\CN=4$ superfields have many fermionic components which we need
to reduce in number by a constraint. At the same time we wish to
recover the transformations \extenansa. Both of these goals are
accomplished by the constraint
\eqn\con{\eqalign{
\cD_i {\bf \Phi}^{Aj} &= \epsilon_i{}^j{}_k \cD_{}{\bf \Phi}^{Ak}
+ \delta^j_i {\bf \Psi}^A
\cr
\cD_i{\bf  \Psi}^A &= -i \delta_{ij} \dot{{\bf \Phi}}^{Aj}.
}}
One recovers \extenansa\ by plugging these constraints into the
$\theta_i$-independent part of \extendedtrans.

A manifestly $\CN =4$ supersymmetric action can now be constructed as
\eqn\s{
S = {1 \over 2} \int dt d^4\theta \ L({\bf \Phi}).
}
This can be reduced to an $\CN=1$ superspace action using the constraints
\con\ and integrating over the three $\theta_i$.
Using $\int d\theta_i {\bf F} = \left. \cD_i {\bf F} \right|_{\theta_i=0}  
+ (total \ derivative)$
we find that
\eqn\sl{\eqalign{
S = {1 \over 2} \int dt d\theta_{}\
 \Big[ &L_{,\mu\nu\rho} \cD_1{\bf \Phi}^\mu \cD_2{\bf \Phi}^\nu
\cD_3{\bf \Phi}^\rho
\cr
&\qquad
+ L_{,\mu\nu} ( \cD_1{\bf \Phi}^\mu
\cD_2 \cD_3 {\bf \Phi}^\nu + \cD_2{\bf \Phi}^\mu \cD_3 \cD_1 {\bf \Phi}^\nu
            + \cD_3{\bf \Phi}^\mu \cD_1 \cD_2 {\bf \Phi}^\nu)
\cr
&\qquad+ L_{,\mu} \cD_1 \cD_2 \cD_3 {\bf \Phi}^\mu
\Big]_{\theta_i=0}
}}
where $\mu,\nu,\rho = 1,\dots,3N$ run over the moduli space indices
$Aj$.  From \con\ we see that for $k \ne i$
\eqn\ds{
\cD_k \cD_i {\bf \Phi}^{Aj} = - \cD_i \cD_k {\bf \Phi}^{Aj} =
-\epsilon_i{}^j{}_k \cD_{} {\bf \Psi}^A
- 2 i \delta_{m [ k} \delta_{i]}{}^j \dot{\bf \Phi}^{Am},
}
\eqn\dss{
\cD_1 \cD_2 \cD_3 {\bf \Phi}^\mu = - i \cD_{} \dot{\bf \Phi}^\mu
.}
We may write the action in a more symmetric form by anticommuting and
integrating the third line of \sl\ by parts:
\eqn\sll{\eqalign{
S = {1 \over 2}
 \int dt d\theta_{} \ \bigg[ & {1 \over 3!} L_{,\mu\nu\rho} \epsilon^{ijk}
            \cD_i{\bf \Phi}^\mu  \cD_j{\bf \Phi}^\nu
\cD_k{\bf \Phi}^\rho
\cr
&\qquad +
L_{,\mu\nu} ( {\textstyle{1 \over 2}}
\epsilon^{ijk} \cD_i {\bf \Phi^\mu} \cD_j \cD_k {\bf \Phi^\nu}
            + i \cD_{} {\bf \Phi^\mu }{\bf \dot{\Phi}^\nu })
\bigg]_{\theta_i=0}.}}
The most general supersymmetric action for $\Phi^\mu$
and $\Psi^A$ contains the terms \coles
\eqn\sq{\eqalign{
S = \int dt d\theta_{}\  \bigg[ &{i \over 2} g_{\mu\nu}
\cD\Phi^\mu \dot{\Phi}^\nu
- {1 \over 2} h_{AB} \Psi^A \cD\Psi^B - i f_{\mu A} \dot{\Phi}^\mu \Psi^A
+ {1 \over 3!} c_{\mu\nu\rho} \cD\Phi^\mu \cD\Phi^\nu \cD\Phi^\rho \cr
&\qquad + {1 \over 2!} n_{\mu\nu A} \cD\Phi^\mu \cD\Phi^\nu \Psi^A
+ {1 \over 2!} m_{\mu A B} \cD\Phi^\mu \Psi^A \Psi^B
+ {1 \over 3!} l_{ABC} \Psi^A \Psi^B \Psi^C
\bigg]
.}}
Using \con\ and \ds\ we can read off the quadratic terms of \sll,
\eqn\g{
g_{Ai Bj} = (\delta^k_i \delta^l_j +
    \epsilon^{mk}{}_i \epsilon_m{}^l{}_j
)\ \p_{Ak} \p_{Bl} L,
}
\eqn\h{
h_{AB} = \delta^{ij} \p_{Ai} \p_{Bj} L,
}
\eqn\f{
f_{Ai B} = \epsilon^{jk}{}_{i} \p_{Aj} \p_{Bk} L
.}
The $\cD\Psi \cD\Phi$ term has been
integrated by parts and absorbed into the $f$ and $n$ terms.
The cubic terms are
\eqn\c{
c_{Ai Bj Ck} = {1 \over 2} \epsilon^{pqh}
\epsilon_p{}^l{}_i \epsilon_q{}^m{}_j \epsilon_h{}^n{}_k
\
\p_{Al} \p_{Bm} \p_{Cn} L,
}
\eqn\n{
n_{Ai Bj C} = \half (  \epsilon^{pqn} \epsilon_p{}^l{}_i
\epsilon_q{}^m{}_j + \epsilon^l{}_i{}^n\delta^m_j
- \epsilon^m{}_j{}^n\delta^l_i)
\ \p_{Al} \p_{Bm} \p_{Cn} L,
}
\eqn\m{
m_{Ai B C} = {1 \over 2} \epsilon^{jmn}\epsilon_j{}^l{}_i
            \ \p_{Al} \p_{Bm} \p_{Cn} L,
}
\eqn\lterm{
l_{ABC} = {1 \over 2 } \epsilon^{lmn} \ \p_{Al} \p_{Bm} \p_{Cn} L
.}

The actions \s\ and \sq\ are $\CN=4$ supersymmetric for any
function $L$.  Comparing the bosonic metric \g\ appearing in the
action with the moduli space metric \modmetric\ we conclude that
the choice
\eqn\choice{L=-{1 \over 16 \pi}\int d^3x\ \psi^4}
describes the $\CN=4$ supersymmetric quantum mechanics of $N$ black holes.

It is straightforward to check that with $L$ given by
\choice, the coupling $f_{\mu A}$ computed from \f\ is nonzero (as long
as $N>2$).  From
\sq\ it can be seen that this implies a coupling of the form
$f_{\mu A} \dot{X}^\mu b^A$ once the superfields are written out
in terms of components.  When the auxiliary fields $b^A$ are integrated
out, the quadratic action $g_{\mu \nu} \dot{X}^\mu \dot{X}^\nu$
will receive an additional contribution proportional to
$f_{\mu A} f_{\nu B} h^{AB} \dot{X}^\mu \dot{X}^\nu$.
This signifies a modification of the moduli space metric required
by supersymmetry.

\newsec{Superconformal Symmetry with Fermion Multiplets}

In this section we investigate the superconformal extension of the
supersymmetry algebra developed in the previous section.  We continue
to work with the multiplets $(X^\mu,\lambda^\mu)$ and $(\psi^A,b^A)$.
In sections 4.1 and 4.2 we will work with generic
$I_i$ and $e_i$, requiring only that the extra supersymmetries
\extenansa\ satisfy the supersymmetry algebra \susyalgebra\ -- we
will restrict our attention to the specific choices of
$I_i$ and $e_i$ \ourie\ only in section 4.3.
Although our approach
will resemble that of \sqm\ we will work entirely in the lagrangian
formulation.  One consequence of this is that we will investigate
separately the closure of the superconformal algebra on the fields
$(X^\mu,\lambda^\mu,\psi^A,b^A)$ and invariance of the action \sl.

\subsec{Conformal Transformations}

We investigate the behavior of the fields
$(X^\mu,\lambda^\mu,\psi^A,b^A)$ under conformal transformations.
It is convenient to parametrize a conformal transformation by
\eqn\confparm{
\epsilon(t) = \epsilon_H + 2 t \epsilon_D + t^2 \epsilon_K.
}
where $\epsilon_H$, $\epsilon_D$, and $\epsilon_K$ are constant
infinitesimal parameters corresponding respectively to time
translations, dilatations and special conformal transformations.
With this parametrization the $SL(2,{\bf R})$ algebra takes the form
\eqn\slrone{
[\delta_{\epsilon_1}, \delta_{\epsilon_2}] = \delta_{[\epsilon_1,
\epsilon_2]}, \qquad {\rm where}~[\epsilon_1,
\epsilon_2] = \epsilon_1 \dot{\epsilon}_2 - \epsilon_2 \dot{\epsilon}_1.
}
If we define the generators $H$, $D$ and $K$ by
\eqn\hdk{
\delta_{\epsilon_H} = i \epsilon_H H, \qquad
\delta_{2 \epsilon_D t} = i \epsilon_D D, \qquad
\delta_{\epsilon_K t^2} = i \epsilon_K K,
}
then the algebra takes the familiar form
\eqn\slr{
[ H, K] = - i D, \qquad [ H, D] = -2 i H, \qquad [K, D] = 2 i K.
}
The variation of the field $X^\mu$ under a conformal transformation
is
\eqn\confx{
\delta_\epsilon X^\mu = - \epsilon \dot{X}^\mu + \half \dot{\epsilon} D^\mu
}
for some vector field $D^\mu(X)$.
One easily checks that the $SL(2,{\bf R})$ algebra
\slrone\ is satisfied for any $D^\mu$.
For the remaining fields, we make the
ans\"atze\foot{Although
we shall not find it necessary, one could consider more
general conformal transformations
where $D^\mu$, $F^\mu$, $G^A$ and $H^A$ are functions of
$(X^\mu,\lambda^\mu,\psi^A,b^A)$.
In this case the conditions required by closure of the $Osp(1|2)_r$
algebras become somewhat more complicated.
}
\eqn\ansatze{\eqalign{
\delta_\epsilon \lambda^\mu &= - \epsilon \dot{\lambda}^\mu + \half
\dot{\epsilon} F^\mu,\cr
\delta_\epsilon \psi^A &= - \epsilon \dot{\psi}^A + \half \dot{\epsilon}
G^A,\cr
\delta_\epsilon b^A &= - \epsilon \dot{b}^A + \half \dot{\epsilon} H^A.
}}
Again one easily checks that the $SL(2,{\bf R})$ algebra is
satisfied as long as $F^\mu(X,\lambda)$, $G^A(X,\psi)$ and
$H^A(X,b)$ do not depend on the time derivatives of the four basic
fields.

We now wish to enlarge the
algebra to include the supersymmetries
\qdef\ and \extenansa. 
In analogy with the above discussion we will express the supersymmetry
variations in terms of generators $Q^r$
of supersymmetry transformations on the component fields
(not to be confused with 
the superderivatives \ddef, which act on superfields) defined by
\eqn\gendef{
\delta_{\zeta} = i \zeta_r Q^r
}
where the $\zeta_r$ are anticommuting parameters.
Since the supersymmetry transformations do not involve the index $A$
we will suppress this index and consider
just the fields $(X^i,\lambda^i,\psi,b)$.
The 
action of these generators
on the component fields is
\eqn\susyonecomp{
Q^0 X^i =-\lambda^i, \qquad Q^0 \lambda^i = -i \dot{X}^i, \qquad
Q^0 \psi = -i b, \qquad Q^0 b =- \dot{\psi}
}
and
\eqn\extendedcomp{\eqalign{
Q^i X^j &= - \epsilon^{ij}{}_k \lambda^k + \delta^{ij} \psi,\cr
Q^i \lambda^j &= i \epsilon^{ij}{}_k \dot{X}^k - i \delta^{ij} b,\cr
Q^i \psi &=  i\dot{X}^i,\cr
Q^i b &= -  \dot{\lambda}^i.
}}
It is straightforward to verify that these generators 
obey the required anticommutation relations $\{Q^r,Q^s\} = 2 \delta^{rs}H$.

We first enlarge the $SL(2,{\bf R})$
algebra to $Osp(1|2)_0$ by incorporating
the $\CN=1$ supersymmetry transformation $Q^0$.
Following the procedure outlined in Appendix A
we define the superconformal generator $S^0$ by the relation
$[K,Q^0] = i S^0$.
The requirement that $[D,Q^0]=i Q^0$ on $X^\mu$ implies the unique
choice $F^\mu = Q^0(X^\mu - D^\mu)$.
This same relation on $\psi^A$ requires
$H^A = i Q^0 (G^A - \psi^A)$.
The relation $[K, S^0] = 0$ gives no additional constraints, so
the
remaining (anti-) commutation relations of $Osp(1|2)_0$ (see Appendix
B) follow with
no additional restrictions
on $D^\mu(X)$ or $G^A(X,\psi)$.

The next step is to incorporate the extended supersymmetries \extendedcomp.
We first examine the conditions required by closure of the $Osp(1|2)_i$
subalgebra, which is generated by $(H,D,K,Q^i,S^i)$, for each value of $i$.
We define $S^i$ by $[K,Q^i] = i S^i$ as before.
We find
\eqn\dqi{
[D, Q^i] X^\mu = i Q^i X^\mu + i (\CL_D {I^{i\mu}}_\nu) \lambda^\nu
-  \psi^A ( \CL_D {e^{i\mu}}_A + {e^{i\mu}}_A) - {e^{i\mu}}_A G^A.
}
so that closure of the algebra requires $\CL_D I_i{}^\mu{}_\nu = 0$.
Furthermore, if we make the ansatz that $(\CL_D - \beta) e_i{}^\mu{}_A = 0$
for some constant $\beta$ then \dqi\ requires
\eqn\aaa{
G^A = - (\beta+1) \psi^A.
}
Acting on $\psi^A$ we find
\eqn\aaa{
[D, Q^i] \psi^A = i Q^i \psi^A +  \dot{X}^\mu
(\CL_D + \beta) e_i{}^A{}_\mu
}
so we must have
\eqn\betadef{
(\CL_D + \beta) e_i{}^A{}_\mu = 0.
}
The rest of the $Osp(1|2)_i$ algebra follows without further
restrictions.

We have verified that the supersymmetry transformations
\qdef\ and \extenansa\ together with the conformal transformations
\eqn\conf{\eqalign{
\delta_\epsilon X^\mu &= - \epsilon \dot{X}^\mu + \half \dot{\epsilon} D^\mu,\cr
\delta_\epsilon \lambda^\mu &= - \epsilon \dot{\lambda}^\mu + \half
\dot{\epsilon} ( {D^\mu}_{,\nu} \lambda^\nu - \lambda^\mu),\cr
\delta_\epsilon \psi^A &= - \epsilon \dot{\psi}^A - \half \dot{\epsilon}
(\beta+1) \psi^A,\cr
\delta_\epsilon b^A &= - \epsilon \dot{b}^A - \half\dot{\epsilon}
(\beta+2) b^A
}}
satisfy the $\CN$ separate algebras $Osp(1|2)_r$ as long as
\eqn\conditions{
\CL_D I_i{}^\mu{}_\nu=
(\CL_D - \beta) e_i{}^\mu{}_A
= (\CL_D + \beta) e_i{}^A{}_\mu = 0
}
for each $i = 1,\ldots,\CN-1$.

We must now knit these $\CN$ $Osp(1|2)_r$ algebras together into the
appropriate superalgebra.  Following the procedure outlined in
Appendix A, it remains only to check that the $Q^r$ lie in an
appropriate spinor representation of the $R$ symmetry algebra
which appears on the right-hand side of the $\{Q^r,S^s\}$ anticommutator.
This will be done for the constant $I$ and $e$ 
case \ourie\ in section 4.3 below.

\subsec{Conformally Invariant $\CN=1$ Actions}

In this section we construct conformally invariant actions
out of the $\CN=1$ multiplets
$\Phi^\mu$ and $\Psi^A$.  Let us start with the superfield $\Phi^\mu$.
The most general action
involving only dimensionless couplings is \coles
\eqn\actionone{
S_1 = \int dt d\theta\ {i \over 2} g_{\mu\nu} \cD \Phi^\mu
\dot{\Phi}^\nu  + {1 \over 6} c_{\mu\nu\rho}
\cD \Phi^\mu \cD \Phi^\nu \cD \Phi^\rho.
}
In terms of component fields
\eqn\sone{
S_1 =
 \int dt \ {1 \over 2} g_{\mu\nu}
[ \dot{X}^\mu \dot{X}^\nu + i \lambda^\mu D_t \lambda^\nu]
+ {i \over 2} c_{\mu\nu\rho} \lambda^\mu \lambda^\nu \dot{X}^\rho
- {1 \over 6} c_{\mu\nu\rho,\sigma}
\lambda^\mu\lambda^\nu\lambda^\rho\lambda^\sigma,
}
where
\eqn\aaa{
D_t \lambda^\nu = \dot{\lambda}^\nu + {\Gamma^\nu}_{\rho\sigma}
\dot{X}^\rho \lambda^\sigma.
}
It is convenient to consider separately the bosonic and fermionic
terms in \sone.  Using \conf\ we find
\eqn\aaa{
\delta_\epsilon \left(
\half  g_{\mu\nu} \dot{X}^\mu \dot{X}^\nu
\right)\sim
{1 \over 4}
\dot{\epsilon} [ (\CL_D - 2) g_{\mu\nu}] \dot{X}^\mu
\dot{X}^\nu + {1 \over 2} \ddot{\epsilon} D_\mu \dot{X}^\mu,
}
where $\sim$ denotes equality up to total
derivatives.
Thus invariance of this term under dilatations requires
\eqn\condone{
(\CL_D -2) g_{\mu\nu} = 0,
}
i.e. that $D^\mu$ is a homothety.
Invariance under special conformal transformations
requires in addition that
\eqn\exact{
D_\mu = \p_\mu K
}
for some function
$K$.
A vector field 
$D^\mu$ obeying \condone\ and \exact\ is known as a closed homothety.
Note that in general invariance of the action under
dilatations does not guarantee invariance under the full
conformal group.
In all further calculations we assume \condone\ and \exact\ hold.
For the two-fermion terms in \sone\ we find
\eqn\aaa{
\delta_\epsilon \left( \itwo g_{\mu\nu}
\lambda^\mu D_t \lambda^\nu + \itwo
c_{\mu\nu\rho} \lambda^\mu \lambda^\nu \dot{X}^\rho \right) \sim
{i \over 4} \dot{\epsilon} [ (\CL_D - 2) c_{\mu\nu\rho}]
\lambda^\mu \lambda^\nu \dot{X}^\rho + {i \over 4}
\ddot{\epsilon}
D^\mu c_{\mu\nu\rho} \lambda^\nu \lambda^\rho
}
so invariance requires that
\eqn\condtwo{
(\CL_D -2) c_{\mu\nu\rho} = 0, \qquad D^\mu c_{\mu\nu\rho} = 0.
}
Again, the first condition is required by dilatation invariance
and the second is an additional constraint
required for full conformal symmetry.
Finally, varying the four-fermion terms in \sone\ gives
\eqn\aaa{
\delta_\epsilon \left(
-{\textstyle{1 \over 6}} c_{\mu\nu\rho,\sigma}
\lambda^\mu\lambda^\nu\lambda^\rho\lambda^\sigma
\right)  \sim
{1 \over 12} \dot{\epsilon} \p_\sigma [ (\CL_D - 2) c_{\mu\nu\rho}]
\lambda^\sigma\lambda^\mu\lambda^\nu\lambda^\rho,
}
which vanishes as consequence of \condtwo, giving no further constraints.
The conditions \condone, \exact\ and \condtwo\ agree precisely with those
found by the authors of  \sqm, who used the Hamiltonian formalism.

When the fermion multiplet $\Psi^A = i \psi^A
+i  \theta b^A$ is included there are five additional
terms that one can add to the superspace
lagrangian \actionone,
\eqn\newterms{\eqalign{
\CL_2 &= -{1 \over 2}  h_{AB} \Psi^A \cD \Psi^B
+ {1 \over 6}  l_{ABC} \Psi^A \Psi^B \Psi^C
\cr
&\qquad\qquad\qquad\qquad
-i f_{\mu A} \dot{\Phi}^\mu \Psi^A
 +{1 \over 2}  m_{\mu A B} \cD \Phi^\mu \Psi^A \Psi^B
 + {1 \over 2}  n_{\mu\nu A} \cD \Phi^\mu
\cD \Phi^\nu \Psi^A.
}}
The calculation of $\delta_{\epsilon} \CL_2$ is similar
to the above calculation, so we will simply quote the result.
The terms \newterms\ are dilatation invariant provided that
\eqn\dilationfermion{\eqalign{
\CL_D h_{AB} &= (2 \beta + 2) h_{AB},\cr
\CL_D l_{ABC} &= (3 \beta + 2) l_{ABC},
\cr
\CL_D f_{\mu A} &= (\beta + 2) f_{\mu A},\cr
\CL_D m_{\mu A B} &= (2 \beta + 2) m_{\mu A B},\cr
\CL_D n_{\mu \nu A} &= (\beta+2) n_{\mu \nu A}, }}
and invariant under special conformal transformations if, in addition,
\eqn\conformalfermion{\eqalign{
0 &= D^\mu m_{\mu A B},\cr
0 &= D^\nu ( n_{\mu \nu A} - \nabla_\mu f_{\nu A}).
}}
Note that the $n_{\mu \nu A}$ and $f_{\mu A}$ terms in \newterms\ mix under
conformal transformations.

\subsec{$D(2,1;\alpha)$ Quantum Mechanics with Fermion Multiplets}

In this section we work out the $R$-symmetries and full
superconformal algebra for the special case \ourie\ with constant
$I$ and $e$.
We assume the existence of a closed homothety of the form
\eqn\gauge{
D^{Ai} = {2 \over h} X^{Ai}
}
for some constant $h$.
Since $e_i{}^A{}_\mu$ is now $e_i{}{}^A{}_{B j} = \delta^A_B \delta_{ij}$,
which is constant, we have
\eqn\aaa{
\CL_D e_i{}^A{}_\mu = e_i{}^A{}_\nu D^\nu{}_{,\mu} =
{{2 \over h}} e_i{}^A{}_\mu.
}
Comparing with \betadef\ we see that $\beta$ and $h$ must be related
by $\beta = -{2 \over h}$.

The first step is to find the superconformal generators $S^r$,
which are defined by $S^r = i [ Q^r, K]$.  Using
\susyonecomp\ and \extendedcomp, and again suppressing the $A$
index, we find
\eqn\aaa{
\eqalign{
S^0 X^j &= - t \lambda^j, ~~~~\!~~~~~~~~~~~~~
S^i X^j = - t \epsilon^{ij}{}_k \lambda^k + t \delta^{ij} \psi,\cr
S^0 \lambda^j &= -i t \dot{X}^j + {\textstyle{2 i \over h}} X^j,
~~~~~
S^i \lambda^j = i t \epsilon^{ij}{}_k \dot{X}^k - i t \delta^{ij} b
- {\textstyle{2 i \over h}}
\epsilon^{ij}{}_k X^k,\cr
S^0 \psi &= - i t b, ~~~~~~~~~~~~~~~~~~~
S^i \psi =  i t \dot{X}^i - {\textstyle{2 i \over h}} X^i,
\cr
S^0 b &= - t \dot{\psi} + ({\textstyle{2 \over h}}-1) \psi,
~~~~~
S^i b = - t \dot{\lambda}^i + \left({\textstyle{2 \over h}}-1\right)
\lambda^i.
}}
As per the discussion in Appendix A, the hard part is now to
package the $\{Q^r, S^s\}$ anticommutator into a nice form by
defining the appropriate $R$ symmetry generators.
We find that the $\{Q,S\}$ anticommutator has the form
\eqn\dalgebra{
\{ Q^r, S^s \} = D \delta^{rs} + {4 (h+1) \over h}
(T_{+i}){}^{rs}R^i_+ - {4 \over h} (T_{-i}){}^{r s} R^i_-,
}
where the $T_\pm^i$ are constant matrices defined by
\eqn\tdef{
(T_\pm^i){}^{rs}
= \mp \delta^{i[r} \delta^{s]0} + {\textstyle{1 \over 2}} \epsilon^{irs},
}
and the $R$ symmetry generators are given by
\eqn\aaa{\eqalign{
R_\pm^i X^j &= {\textstyle{i \over 2}} (1 \mp 1) \epsilon^{ij}{}_k X^k,\cr
R_\pm^i \lambda^j &= {\textstyle{i \over 2}} (\epsilon^{ij}{}_k \lambda^k \mp
\delta^{ij} \psi),\cr
R_\pm^i \psi &= \pm {\textstyle{i \over 2}} \lambda^i,\cr
R_\pm^i b &= 0.
}}
The $T_\pm^i$ satisfy
\eqn\trelations{
[ T_+^i, T_-^j ] = 0, \qquad
[ T_\pm^i, T_\pm^j ] = - \epsilon^{ij}{}_k T_\pm^k,
\qquad
\{ T_\pm^i, T_\pm^j \}
= - {1 \over 2} \delta^{ij},
}
and the $R$ symmetries satisfy
\eqn\aaa{
[ R_+^i, R_-^j ]  =0, \qquad
[ R_{\pm}^i, R_{\pm}^j ] = i \epsilon^{ij}{}_k R_\pm^k.
}
Thus the $R$-symmetry of this theory is $SU(2) \times SU(2)$.
The $R_-$ act on the $X^{Aj}$ as an $SO(3)$ triplet,
so we interpret this $SU(2)$ symmetry as arising from the $SO(3)$ 
spatial rotations of the original theory.\foot{A more careful analysis
shows that $SO(3)$ spatial rotations are generated by the diagonal
subgroup $R_- + R_+$.}
The $X^{Ai}$ are uncharged under $R_+$, so the second $SU(2)$ must
come from the $SU(2)$ $R$-symmetry of the original $D=4$,
$\CN=2$ supergravity.
The four $Q$'s transform as complex doublets of each $SU(2)$,
\eqn\rq{
[ R_\pm^i, Q^0 ] = \pm {i \over 2} Q^i, \qquad
[R_\pm^i, Q^j ] = {i \over 2} ( \mp \delta^{ij} Q^0 +
\epsilon^{ij}{}_k Q^k ),
}
and the $S$'s similarly.
We recognize \dalgebra\ and \rq\ as the defining relations of the
$D(2,1;\alpha)$ superalgebra with parameter
\eqn\alphadef{
\alpha = - h - 1.}

In conclusion, we have demonstrated that an $\CN=4$ supersymmetric theory 
with action \s\ 
has $D(2,1;\alpha=-h-1)$ symmetry if it admits a closed 
homothety of the form \gauge. 

\newsec{Superconformal Symmetry of the Near-Horizon Moduli Space }

In this section we demonstrate that the quantum mechanics defined
by \modmetric\ admits a $D(2,1;0)$  symmetry in the
near-horizon limit. In the near-horizon limit the metric is 
\eqn\gform{ g_{Ak Bl} = G^{ij}_{kl} \p_{Ai} \p_{Bj} L } where
\eqn\gijkl{ G^{ij}_{kl} = \delta^i_k \delta^j_l +
\epsilon^{mi}{}_k \epsilon_m{}^j{}_l } and \eqn\l{ L (x^{Ai}) = -
{1 \over 16 \pi} \int d^3x \ \left[\sum_A { m_A \over | {\bf x} -
{\bf x}^A|}\right]^4. } Our ansatz for the homothety is \eqn\aaa{
D^{Ai} = {2 \over h} x^{Ai} } for some constant $h$. It follows from 
\gform\ and \gijkl\ 
that $D$ is a homothety \condone\ if  \eqn\dh{ (x^{Ai}\p_{Ai}
- h) L = {h \over 2} K } where at this point $K$ can be any
function in the kernel of the differential operator of \gform ,
i.e. \eqn\phi{ G^{ij}_{kl} \p_{Ai} \p_{Bj} K = 0 .}

We should be careful since
$L$ contains a divergent piece that does not contribute to the metric.
To see this let us separate out two terms as
\eqn\li{
L = L_1 + L_2 + L'
}
where
\eqn\lone{
L_1 =
- {1 \over 16 \pi} \sum_A m_A^4 \int d^3x \ { 1 \over | {\bf x} - {\bf x}^A|^4}
}
and
\eqn\ltwo{
L_2 =
- {1 \over 4 \pi} \sum_{A \ne B} m_A^3 m_B \int d^3x \
{ 1 \over  |{\bf x} - {\bf x}^A|^3 |{\bf x} - {\bf x}^B|}
.}
These are the only two potentially divergent pieces since all other
terms in $L$ contain at most an integrable singularity
$|{\bf x}-{\bf x}^A|^{-2}$
as ${\bf x} \to {\bf x}^A$ (provided that none of
the black holes are coincident).
Moreover, $L'$ is homogeneous of degree $-1$, i.e.
$L' (\lambda x^{Ai}) = \lambda^{-1} L'( x^{Ai})$,
so Euler's theorem tells us that
\eqn\lpd{
(x^{Ai} \p_{Ai} + 1 )L' = 0
.}
Let us now turn our attention to the divergent terms.
First, note that $L_1$ is
independent of ${\bf x}^A$ and thus does not contribute to the metric.
If we insert a cutoff $|{\bf x} - {\bf x}^A| > \delta$ in the integral
\ltwo\ we find that
\eqn\cut{
L_2 = - \sum_{A \ne B} m_A^3 m_B
{ \ln r_{AB} + (1 - \ln \delta) \over r_{AB}}.
}
However, using
$G^{(ij)}_{kl} = \delta^{ij} \delta_{kl}$
it is easy to show that
\eqn\zero{
G^{ij}_{kl} \p_{Ci} \p_{Dj} {1 \over r_{AB}} = 0
}
so that the $1- \ln \delta$ term does not contribute to the metric.
Thus we find that $L_2$ is {\it not} homogeneous but instead
satisfies
\eqn\dl{
(x^{Ai} \p_{Ai} + 1) L_2 = -{1 \over 2 } K
}
where
\eqn\k{
K= 2 \sum_{A \ne B} m_A^3 m_B {1 \over r_{AB}}
}
is in the kernel of $G^{ij}_{kl}\p_{Ci}\p_{Dj}$ by \zero.
So \dh\ holds for $h=-1$ and
\eqn\d{
D^{Ai} = -2 x^{Ai}
}
is a homothetic vector field.

We now show that \exact\ holds, i.e. that
\eqn\gcond{
D_{Ak} = g_{AkBl} D^{Bl} = - 2 g_{AkBl} x^{Bl} = \p_{Ak} K
.}
It turns out (as anticipated by the notation) that the solution to this
is precisely the function $K$
defined in \k. To see this, note that
$L$ obeys $L(O^i_j x^{Aj}) = L(x^{Ai})$ for any orthogonal matrix
$O^i_j$.  Thus
\eqn\neweqn{
x^{Bk} \epsilon^{mi}{}_k \p_{Bi} L = 0
} since the $\epsilon^{mi}{}_k$ generate the $SO(3)$ symmetry of
${\bf R}^3$. A straightforward computation then reveals that
\gcond\ holds. This completes the proof that the bosonic part of
the sigma model action admits an $SL(2,{\bf R})$ symmetry in the
near-horizon limit.

It remains to show that the fermionic terms arising from the supersymmetric
completion of the action are also conformally invariant.
It is straightforward to verify from \d, \neweqn\ and \h--\lterm\ that
conditions \condtwo, \dilationfermion\ and \conformalfermion\
hold for $\beta = 2$. Since we have $h=-1$ the superconformal
group is $D(2,1;0)$. This is a semi-direct
product of $SU(1,1|2)$ with
$SU(2)$, where the extra $SU(2)$ acts nontrivially on the supercharges
of $SU(1,1|2)$.

The generator $K$ of special conformal transformations has some 
useful features.
Conformal invariance in quantum mechanics was 
studied in 
\refs{\ddf,\jackiw}
(following the more general treatment of \refs{\cj,\ccj}), wherein 
it is noted that the hamiltonian $H$ of such a theory 
possesses neither a ground state nor discrete eigenstates.
It was suggested that one should consider, instead of $H$
eigenstates, eigenstates of
\eqn\lnought{
L_0 = {1 \over 2}(H+K),
}
which has a well behaved discrete spectrum of normalizable
eigenstates.
In our case, the
near-horizon limit of
the black hole moduli space has
asymptotically locally flat
regions corresponding to near-coincident black holes
$r_{AB} \to 0$ for any $A \ne B$, so the hamiltonian does
not have a ground state or discrete eigenstates. 
However, the function $K$ \k\ diverges in
these noncompact regions of the moduli
space.  Thus the operator $L_0$, in which $K$ serves as
a potential to cut off the noncompact regions of the moduli space,
will provide a sensible definition of the quantum mechanics,
as per the suggestion of \ddf\ (and in a related  black hole context of 
\refs{\claus, \gt}).  A similar story was recently
found for five-dimensional black holes \refs{\fived,\iceland,\twoBH}.

\vskip .7cm

\centerline{\bf Acknowledgements}

\vskip .2cm

It is a pleasure to thank
R. Britto-Pacumio, G. Papadopoulos, M. Headrick,
R. Jackiw,
A. Volovich and especially J. Michelson
for many helpful conversations.
This work was supported
by the NSF graduate fellowship program and DOE grant DE-FG02-91ER40654.

\appendix{A}{A Construction of $d = 1$ Superconformal Algebras}

A generic superconformal algebra in $d=1$ dimensions contains
the conformal group $SL(2,{\bf R})$ generated by $H$,
$D$ and $K$, $\CN$ fermionic supercharges $Q^i$, an equal number of
fermionic partners $S^i = i [Q^i, K]$, as well as some number of
bosonic $R$-symmetry generators required for closure of the algebra.
A classification of the possible $d=1$ superconformal algebras is
obtained by reading off from Nahm's classification
\nahm\ of superalgebras those in which $SL(2,{\bf R})$ is a
factored subgroup of the bosonic part of the superalgebra
and in which the fermionic generators sit in a spinorial representation
of $SL(2,{\bf R)}$ \kallosh.  The result appears in Table 1.

\bigskip

{\centerline{
\vbox{\offinterlineskip
\hrule
\halign{&\vrule#&
\strut\quad\hfil#\quad\cr
height2pt&\omit&&\omit&&\omit&&\omit&\cr
&\hfill Superalgebra\hfill&&\hfill $dim (\#b, \#f)$
\hfill&& \hfill $R$-symmetry
\hfill &&\hfill Spinor Rep.\hfill &\cr
height2pt&\omit&&\omit&&\omit&&\omit&\cr
\noalign{\hrule}
height2pt&\omit&&\omit&&\omit&&\omit&\cr
& \hfill $Osp(1|2)$ \hfill && \hfill (3,2) \hfill && \hfill $1$ \hfill 
&& \hfill {\bf 1} \hfill & \cr
\noalign{\hrule}
height2pt&\omit&&\omit&&\omit&&\omit&\cr
& \hfill $Osp(2|2) = SU(1,1|1)$
\hfill && \hfill (4,4) \hfill && \hfill $U(1)$ \hfill 
&& \hfill {\bf 2} \hfill & \cr
\noalign{\hrule}
height2pt&\omit&&\omit&&\omit&&\omit&\cr
& \hfill $Osp(3|2)$ \hfill && \hfill (6,6) \hfill && \hfill $SU(2)$ \hfill
&&\hfill {\bf 3} \hfill
& \cr
\noalign{\hrule}
height2pt&\omit&&\omit&&\omit&&\omit&\cr
& \hfill $PSU(1,1|2)$ \hfill && \hfill (6,8) \hfill && \hfill $SU(2)$ \hfill
&& \hfill ${\bf 2} \oplus {\bf \bar{2}}$ \hfill & \cr
& \hfill $D(2,1;\alpha)_{0<\alpha\le 1}$
\hfill && \hfill (9,8) \hfill && \hfill $SU(2)\times SU(2)$ \hfill
&& \hfill $({\bf 2},{\bf 2})$ \hfill & \cr
\noalign{\hrule}
height2pt&\omit&&\omit&&\omit&&\omit&\cr
& \hfill $Osp(5|2)$ \hfill && \hfill (13,10) \hfill && \hfill $SO(5)$ \hfill
&& \hfill {\bf 5} \hfill 
& \cr
\noalign{\hrule}
height2pt&\omit&&\omit&&\omit&&\omit&\cr
& \hfill $SU(1,1|3)$ \hfill && \hfill (12,12) \hfill && \hfill $SU(3) \times
U(1)$ \hfill && \hfill ${\bf 3} \oplus {\bf \bar{3}}$ \hfill & \cr
& \hfill $Osp(6|2)$ \hfill && \hfill (18,12) \hfill && \hfill $SO(6)$ \hfill
&& \hfill {\bf 6}\hfill
& \cr
\noalign{\hrule}
height2pt&\omit&&\omit&&\omit&&\omit&\cr
& \hfill $G(3)$ \hfill && \hfill (17,14) \hfill && \hfill $G_2$ \hfill
&& \hfill {\bf 7} \hfill
& \cr
& \hfill $Osp(7|2)$ \hfill && \hfill (24,14) \hfill && \hfill $SO(7)$ \hfill
&&\hfill {\bf 7} \hfill
& \cr
\noalign{\hrule}
height2pt&\omit&&\omit&&\omit&&\omit&\cr
& \hfill $Osp(4^*|4)$ \hfill && \hfill (16,16) \hfill && \hfill $SU(2)
\times SO(5)$ \hfill && \hfill $({\bf 2}, {\bf 4})$ \hfill & \cr
& \hfill $SU(1,1|4)$ \hfill && \hfill (19,16) \hfill && \hfill $SU(4)
\times U(1)$ \hfill && \hfill ${\bf 4} \oplus {\bf \bar{4}}$ \hfill & \cr
& \hfill $F(4)$ \hfill && \hfill (24,16) \hfill && \hfill $SO(7)$ \hfill
&& \hfill ${\bf 8}$ \hfill
& \cr
& \hfill $Osp(8|2)$ \hfill && \hfill (31,16) \hfill && \hfill $SO(8)$ \hfill
&&
\hfill
{\bf 8} \hfill
& \cr
\noalign{\hrule}
height2pt&\omit&&\omit&&\omit&&\omit&\cr
\noalign{\hrule}
height2pt&\omit&&\omit&&\omit&&\omit&\cr
& \hfill $Osp(n|2), n>8$ \hfill && \hfill $(\half n(n-1) + 3,
2 n)$ \hfill && \hfill $SO(n)$ \hfill && \hfill ${\bf n}$ \hfill & \cr
& \hfill $SU(1,1|n), n>4$ \hfill && \hfill $(n^2 + 3, 4n)$ \hfill &&
\hfill $SU(n) \times U(1)$ \hfill && \hfill ${\bf n} \oplus
{\bf \bar{n}}$ \hfill & \cr
& \hfill $Osp(4^*|2n), n>2$ \hfill && \hfill $(2 n^2 + n + 6,8 n)$ \hfill &&
\hfill $SU(2) \times Sp(2n)$ \hfill && \hfill $({\bf 2}, {\bf 2n})$
\hfill & \cr
}
\hrule}
}}

\noindent
{\bf Table 1.}  Lie superalgebras of classical type\foot{An excellent
resource on Lie superalgebras is \dictionary.} that contain
an $SL(2,{\bf R})$ subgroup, adapted from a table in \kallosh.
For clarity we have written out the $\CN\le 8$ algebras
explicitly.

\bigskip

We do not differentiate the various real (i.e, noncompact) forms
of these algebras in our table.  A classification of real simple Lie
algebras of classical type appears in \parker, and the results
also appear in \kallosh.
The algebra
$Osp(4^*|2n)$ has bosonic part $SO^*(4) \times Usp(2n)$, where
$SO^*(4) \cong SL(2,{\bf R}) \times SU(2)$ is a noncompact
form of $SO(4)$.
The superalgebra $PSU(1,1|2)$ is the quotient of $SU(1,1|2)$,
which is not even semi-simple, by the $U(1)$ generated by the
identity matrix.  It has become common in the physics literature
to use $SU(1,1|2)$ as a shorthand for $PSU(1,1|2)$, and we adopt
this convention throughout this paper.

The Lie superalgebras $D(2,1;\alpha)$ with $\alpha \ne 0,-1,\infty$
form a one-parameter family of superalgebras.
The algebras with parameters $\alpha$, $\alpha^{-1}$ and $-1-\alpha$
are isomorphic \dictionary, so it is sufficient to consider
the family of algebras $0 < \alpha \le 1$.  We have
$D(2,1;1) = Osp(4|2) = Osp(4^*|2)$.  In the limit $\alpha
\rightarrow 0$, $D(2,1;\alpha)$ reduces to
a semi-direct product of $SU(1,1|2)$ with $SU(2)$,
with the extra $SU(2)$ acting nontrivially on the fermionic
generators of $SU(1,1|2)$. 

We now describe the construction of a general $d = 1$ superconformal
algebra starting with generators $H$, $D$ and $K$ satisfying the
$SL(2,{\bf R})$ algebra \slr\ and $\CN$ supercharges $Q^i$
satisfying the supersymmetry algebra
\eqn\appaa{
\{ Q^i, Q^j \} = 2 \delta^{ij} H.
}
It follows that
\eqn\appac{
[H, Q^i] = 0.
}
The first nontrivial constraint is that the $Q^i$ must have the
appropriate conformal weight,
\eqn\appab{
[D, Q^i] = i Q^i.
}
We define the superconformal operators $S^i$ through the relation
\eqn\appad{
[K, Q^i] = i S^i.
}
Jacobi identities then guarantee that
\eqn\appadb{
[H, S^i] = -i Q^i, \qquad [D, S^i] = - i S^i,
}
but one must check that the $S^i$ defined by \appad\
satisfy\foot{This is not the only route to take, but it is the one
we find most convenient.}
\eqn\appadc{
[K, S^i] = 0.
}
The definition \appad, together with the appropriate Jacobi identity,
fixes the symmetric part of $\{Q^i, S^j\}$ to be $\delta^{ij} D$.  We
then define the $R$-symmetry generator to be the antisymmetric part,
so that
\eqn\appae{
\{Q^i, S^j\} = \delta^{ij} D + R^{ij}
}
holds.  Note that there are at most $\half \CN (\CN - 1)$ independent
generators $R^{ij}$, so the dimension of the $R$ symmetry algebra
of any $d = 1$ superconformal algebra with $\CN$ supersymmetries can be at
most $\half \CN (\CN - 1)$.  This is just a reflection of the fact
that the $R$ symmetry algebra for $\CN$ real supercharges
can be at most $SO(\CN)$.

Another application of the Jacobi identity gives
\eqn\appaf{
\{S^i, S^j\} - 2 \delta^{ij} K = i [K, R^{ij}].
}
The left-hand side is symmetric under $i\leftrightarrow j$
while the right-hand side is antisymmetric, thus both sides must vanish
separately.  Two more applications of the Jacobi identity give
\eqn\appag{
[H, R^{ij}] = 0 = [D, R^{ij}].
}
We have found from \appaf\ and \appag\ that the $R$
symmetry generators defined by \appae\ commute with $SL(2,{\bf R})$.

We have gotten quite far with little effort, but now it is time to
pay the piper.  As mentioned above, the  $\half \CN (\CN -1)$
generators $R^{ij}$ may or may not
be independent.
We therefore rewrite the $R^{ij}$ in terms of
$dim(\CR)$ independent
generators $R^a$, $a=1, \dots, dim(\CR)$.
The final constraint on the superalgebra is that the $Q^i$
live in a spinor representation of $\CR$, i.e.
\eqn\appah{
[R^a, Q^i] = i (T^a){}^i{}_j Q^j
}
where the $(T^a){}^i{}_j$ satisfy
\eqn\appai{
[T^a, T^b] = - f^{ab}{}_c T^c
}
for some constants $f^{ab}{}_c$.
The $(T^a){}^i{}_j$ are the generators of the representation
and the $f^{ab}{}_c$ are the structure constants of $\CR$.
In general, the requirement \appah\ places very strong constraints
on the theory in question.

We shall now see that \appah\
fixes the rest of the superconformal algebra.
Since $K$ commutes with the $R$
symmetry it follows from \appad\ and \appah\ that
\eqn\appaj{
[R^a, S^i] = i (T^a){}^i{}_j S^j,
}
so the $S^i$ necessarily lie in the same representation of $\CR$ as
the $Q^i$.  Finally, an application of the Jacobi identity to
\appah\ gives
\eqn\appak{
[R^a, R^b] = i f^{ab}{}_c R^c
}
with the help of \appai.  Note that we have reconstructed the algebra
of $\CR$ from the representation \appah, so the representation
must be faithful.

This completes the construction of the superconformal algebra.  In
summary, the construction requires checking \slr, \appaa, \appab,
\appadc, \appah, and \appai. 

\appendix{B}{The $Osp(1|2)$ Algebra}

The ${\cal N} =1 $ superconformal algebra $Osp(1|2)$ contains
$SL(2,{\bf R})$ \slr\ as well as
two fermionic generators $Q$ and $S$
satisfying
\eqn\aaazq{\matrix{
\{ Q, Q \} = 2 H, &&& \{ S, S \} = 2 K, &&& \{ Q, S \} = D,\linesp
[ H, Q ] = 0, &&& [ D, Q] = i Q, &&& [ K, Q ] = i S,\linesp
[ H, S ] = - i Q, &&& [ D, S ] = - i S, &&& [K, S] = 0.
}}

\listrefs

\end